\documentclass[aps,prd,preprint,superscriptaddress,nofootinbib]{revtex4-1}

\usepackage{comment}
\usepackage[utf8]{inputenc} 
\usepackage{graphicx,color,overpic,mathtools}
\usepackage{amsthm,amsmath,amssymb,hyperref,mathrsfs}
\usepackage{braket,bm,bbm,setspace}
\PassOptionsToPackage{normalem}{ulem}
\usepackage{ulem} 
\usepackage{physics}
\usepackage{float}
\usepackage[makeroom]{cancel}
\usepackage[english]{babel}
\usepackage{graphicx}
\graphicspath{ {images/} }
\addto\captionsspanish{}
\hypersetup{
    colorlinks=false,
    pdfborder={0 0 0},
}

\begin{document}

\title{A connection between regular black holes and horizonless ultracompact stars}

\author{Ra\'ul Carballo-Rubio}
\affiliation{CP3-Origins, University of Southern Denmark, Campusvej 55, DK-5230 Odense M, Denmark}
\affiliation{
Florida Space Institute, University of Central Florida, 
 12354 Research Parkway, Partnership 1, 32826 Orlando, FL, USA}
\author{Francesco Di Filippo}
\affiliation{Center for Gravitational Physics, Yukawa Institute for Theoretical Physics, Kyoto University, Kyoto 606-8502, Japan}
\author{Stefano Liberati}
\affiliation{SISSA - International School for Advanced Studies, Via Bonomea 265, 34136 Trieste, Italy}
\affiliation{
IFPU - Institute for Fundamental Physics of the Universe, Via Beirut 2, 34014 Trieste, Italy}
\affiliation{INFN Sezione di Trieste, Via Valerio 2, 34127 Trieste, Italy}
\author{Matt Visser}
\affiliation{
School of Mathematics and Statistics, Victoria University of Wellington, PO Box 600, Wellington 6140, New Zealand
}

\begin{abstract}
We illustrate that regular black holes and horizonless stars, typically considered as quite distinct families of black hole mimickers, are intimately intertwined. We show that any spherically symmetric regular black hole can be continuously deformed into a horizonless star under the mild conditions of non-negativity of gravitational energy (Misner--Sharp quasi-local mass), and an assumed linear relation between the latter and the Arnowitt--Deser--Misner (ADM) mass. We illustrate this general result by considering the family of geometries proposed by Hayward as the description of regular black holes, and we also describe the properties of the corresponding horizonless stars. The form of the associated effective stress-energy tensor shows that these horizonless stars can be identified as anisotropic gravastars with a soft surface and inner/outer light rings. We also construct dynamical geometries that could describe the evolution of regular black holes towards horizonless stars, and show that semiclassical physics contains the necessary ingredients to trigger the early stages of such dynamical evolution.
\end{abstract}


\maketitle

\tableofcontents

\clearpage
\section{Introduction}
 
General Relativity has the peculiar feature that it predicts its own demise via the so called singularity theorems~\cite{Penrose:1964wq,Hawking:1970zqf,Senovilla:2014gza}. Nonetheless such theorems are based, as any mathematical theorem, on a precise set of technical  assumptions which may be partly or completely violated under the extreme conditions reached at the inevitable Planckian densities characterising the near-singularity regime~\cite{Carballo-Rubio:2018jzw,Pandora}. It is widely believed that such singularities are not truly a physical feature of collapsed objects, but rather a red herring, indicating the breakdown of the general relativistic description of spacetime, and that a quantum gravitational treatment is required in order to predict the evolution of the geometry beyond such regions~\cite{Bojowald:2007ky}. 

While it is not at all certain that a continuous and classical description of spacetime would always be possible as a consequence of this quantum gravitational regularization of the singular behaviour, it is not inconceivable that such a  description can represent an adequate proxy almost everywhere for the regularized spacetime. In this sense a vibrant research program probing possible regular spacetimes mimicking general relativistic black holes has been carried out over the years, and has entered in recent times --- mainly thanks to the gravitational wave~\cite{PhysRevLett.116.061102} and shadow observations~\cite{EventHorizonTelescope:2019dse,EventHorizonTelescopeI,EventHorizonTelescopeII,EventHorizonTelescopeIII,EventHorizonTelescopeIV,EventHorizonTelescopeV,EventHorizonTelescopeVI} --- into a revitalized phase of advancement. 

Among the possible scenarios a particular preeminence has been taken by those of regular black holes~\cite{Bardeen1968,Dymnikova:1992ux,Dymnikova:2001fb,Dymnikova:2003vt,Hayward:2005gi,Frolov:2016pav,Carballo-Rubio:2018pmi,Carballo-Rubio:2021bpr,Carballo-Rubio:2022kad} and or horizonless ultracompact stars~\cite{Chapline:2000en,Mazur:2001fv,Mazur:2004fk,Cattoen:2005he,Barcelo:2007yk,Barcelo:2009tpa,Carballo-Rubio:2017tlh,Arrechea:2021xkp}. While the former are solutions characterized by at least an outer and inner horizon but devoid of a singular core, the latter are instead more radical configurations, in which horizons are completely absent but the star surface resides well within the photon sphere of the object, which could have definite phenomenological implications~\cite{Chirenti:2007mk,Cardoso:2019rvt,Carballo-Rubio:2022bgh}.

Regular black hole solutions and horizonless ultracompact stars, have so far been viewed as alternative scenarios for resolving the singular behaviour of spacetime at the core of classical black holes. The purpose of this manuscript is to show that they can instead be viewed as alternative subsets of a given class of regular black-hole-like spacetimes. Note that we use ``stars" instead of the more general term ``objects", as the latter category is broader and includes other geometries such as e.g.~wormholes. It is known that wormholes and regular black holes can coexist within the same family of geometries~\cite{Simpson:2018tsi,Lobo:2020ffi}, which has also recently motivated the study of the associated phenomenology~\cite{Eichhorn:2022oma,Eichhorn:2022fcl,Eichhorn:2022bbn}. However, that this is also true for horizonless stars has not been pointed out before, and both results are not trivially related due to the different topology of the spacetimes involved. The parallelisms and differences between both situations (horizonless stars and wormholes) will be discussed in more detail elsewhere.

First of all, we shall show in Sec.~\ref{sec:RBH} that any family of static non-singular black holes has an associated family of static horizonless stars, while the converse is also true if certain energy conditions are violated by the horizonless stars to be considered. We shall then discuss some of the properties of the horizonless stars associated with the Hayward family of non-singular black holes in Sec.~\ref{eq:haywex}. In Sec.~\ref{sec:dyn}, we study whether these two kinds of black hole mimickers can be related dynamically. These results can be of importance for the understanding of the dynamics of both families of black hole mimickers, providing new clues about formation mechanisms and stability issues. 

\section{Regular black holes and their horizonless compact counterparts}
\label{sec:RBH}

In order to discuss the connection between horizonless compact objects and regular black holes, we shall simplify the discussion by restricting our analysis to spherically symmetric solutions. While of course, rotating/axisymmetric solutions are those of most interest for astrophysical phenomenological studies, the considerations we are going to make herein are easily extendable to these more realistic configurations e.g.~by suitably applying the Newman--Janis ansatz~\cite{newman_note_1965,azreg-ainou_generating_2014, azreg-ainou_static_2014, rajan_complex_2015, rajan_cartesian_2017}
 to the solutions discussed in this work. In what follows, we shall see that for the standard cases of spherically symmetric regular black holes a unified description of the black hole and quasi-black hole limit is possible, with a one-to-one correspondence under suitable assumptions on the matter content of the spacetime.

\subsection{Geometric setting}

As anticipated, in this paper we will be working with spherically symmetric spacetimes. In this case the metric can  without loss of generality be written as
\begin{equation}
\label{eq:spslinel}
    \text{d}s^2=g_{ab}\,\text{d}x^a\,\text{d}x^b=-e^{-2\Phi(v,r)}\left(1-\frac{2m(v,r)}{r}\right)\text{d}v^2
    +2e^{-\Phi(v,r)}\,\text{d}v\,\text{d}r+r^2\text{d}\Omega^2.
\end{equation}
Here, $m(v,r)$ and $\Phi(v,r)$ are arbitrary functions of their arguments, and $\text{d}\Omega^2$ is the line element on the unit 2-sphere. The function $\Phi(v,r)$ must be bounded in the domain of definition of these coordinates for the metric determinant to be well defined. 

The function $m(v,r)$ is the Misner--Sharp--Hernandez (MSH) gravitational energy~\cite{Misner:1964je, Hernandez:1966zia}, also known as the MSH quasi-local mass, which can be invariantly characterized as
\begin{equation}
m(v,r) = \frac{r}{2}\left(1-g^{ab}\,\partial_a r\, \partial _b r\right).   
\end{equation}
Indeed, the right-hand side of the equation above is the definition of the MSH quasi-local energy~\cite{Misner:1964je, Hernandez:1966zia}, as can easily be  checked by noticing that $g^{rr}=1-2m(v,r)/r$. It is also worth mentioning that, as we are working in spherical symmetry, the MSH quasi-local mass is coincident with the Hawking quasi-local energy~\cite{Hawking:1968qt,Hayward:1993ph,Hayward:1994bu}.

It is well known that the MSH energy controls many key properties of the spacetime geometry~\cite{Hayward:1994bu,physical-observability}. In particular, a sphere is trapped if $m(v,r)>r/2$, marginally trapped if $m(v,r)=r/2$ and untrapped if $m(v,r)<r/2$. All information about the structure of (apparent/trapping) horizons is thus encoded in the function $m(v,r)$; in particular, the radial position of trapping horizons $r_{h_i}$ at some given time $v=v_\star$ is implicitly determined by the roots of the equation
\begin{equation}\label{eq:horloc}
 2m(v_\star,r_{h_i})=r_{h_i}.   
\end{equation}
We assume that the metric functions are finite everywhere, which in particular means that near the origin we can write
\begin{align}
    m(v,r)&=m_0(v)+m_1(v)r+m_2(v)r^2+\mathcal{O}(r^3),\nonumber\\
    \Phi(v,r)&=\Phi_0(v)+\Phi_1(v)r+\Phi_2(v)r^2+\mathcal{O}(r^3).
\end{align}
It is then straightforward to conclude \cite{Carballo-Rubio:2019fnb} that, as in the static limit \cite{Frolov:2016pav}, regularity at $r=0$ of the following curvature invariants,
\begin{equation}
    g^{ab}R_{ab},\qquad R_{ab}R^{ab},\qquad R_{abcd}R^{abcd},\qquad C_{abcd}C^{abcd},
\end{equation}
is equivalent to the conditions
\begin{align}\label{eq:metreg}
m_0(v)=m_1(v)=m_2(v)&=0,\nonumber\\
\Phi_1(v)&=0.
\end{align}
These conditions can be equivalently obtained by demanding that the effective energy density and pressures remain finite at $r=0$.

For simplicity, in the following we will drop the time dependence, until Sec.~\ref{sec:dyn} in which we will discuss time-dependent situations in detail. Hence, in the following $m(v,r)$ will have no explicit nor implicit time dependence, and $\Phi(v,r)$ will be an additively separable function $\Phi(v,r) = \Phi_1(r) + \Phi_2(v)$ so that the $v$-dependent part $\Phi_2(v)$ can be reabsorbed in a redefinition of time. It is always possible to reinstate time dependence~\cite{production-and-decay}, with the statements below being still valid for a given instant of time $v=v_\star$. We will also assume the following conditions:
\begin{enumerate}
    \item Asymptotic flatness.
    \item Bounded curvature invariants. 
    \item Non-negative gravitational energy $m(r)$ that displays a parametric dependence on the Arnowitt--Deser--Misner (ADM) mass $M$, as well as an additional length scale $\ell$, \label{cond3}
\begin{equation}\label{eq:stansatz}
m(r)=Mf(r,M,\ell),
\end{equation}
such that there exists a redefinition $\bar \ell = M h(\ell/M)$ for which
    \begin{equation}
    m(r) = M f(r/\bar{\ell}).\label{eq:cond3}
    \end{equation}
That is, we assume that $m(r)$ displays a linear dependence on $M$, plus a generally non-linear dependence on another length scale $\bar{\ell}$. The linearity on $M$ may not be explicit if redefinitions of the parameter $\bar{\ell}$ that involve the former are considered, but it is guaranteed that there exists a parametrization in which the linearity is explicit. We will use both kinds of parametrizations in the discussion below, as changing from one to another is straightforward.

\end{enumerate}

These spacetimes are general enough to describe diverse types of physical situations, in particular regular black holes and horizonless stars. We will first discuss the implementation of these conditions step by step in this section, and then discuss these two situations in detail in Secs.~\ref{sec:rbh} and~\ref{sec:horizonless}, respectively.

The condition of asymptotic flatness is equivalent to
\begin{align}\label{eq:metaflat}
\lim_{r\rightarrow\infty}m(r)&=M,
\nonumber\\
\lim_{r\rightarrow\infty}\Phi(r)&=\Phi_{\infty},
\end{align}
where both constants $M$ and $\Phi_{\infty}$ are finite. The constant $\Phi_{\infty}$ can be absorbed into a rescaling of the null coordinate $v$, and without loss of generality can be set to zero. 

Each of these geometries will be assumed to depend on two parameters, so that we have a family $\{g_{ab}(M,{\ell})\}_{M\in\mathbb{R}^+,{\ell}\in\mathbb{R}^+}$, where $M$ is the ADM mass in its standard definition for asymptotically flat spacetimes and ${\ell}$ is a new length scale.

The position of the trapping horizons for the two metrics $g_{ab}(M,{\ell})$ and $g_{ab}(\lambda M,\lambda{\ell})$ \emph{nominally} change, but \emph{physically} remain the same if we use $\lambda{\ell}$ as the new reference to measure distances and masses. In fact, Eq.~\ref{eq:horloc} can be written (recall that we are dropping the time dependence) as
\begin{equation}\label{eq:horloc1}
 \frac{2M}{\bar{\ell}}\; f(r_{h_i}/\bar{\ell})=\frac{r_{h_i}}{\bar{\ell}}.
\end{equation}
This equation can be written in terms of dimensionless variables $\bar{\alpha} =\bar{\ell} /M\in [0,+\infty)$ and $x =r_{h_i}/\bar{\ell}\in [0,+\infty)$. We will use this form of the equation above later on in our analysis.

Let us now discuss in more detail possible reparametrizations of the length scale $\ell$, namely redefinitions of the length scale $\ell$ so that $\alpha=\ell/M$ changes as $\alpha \rightarrow \bar{\alpha}=\bar{\alpha}(\alpha)$ for a monotonically increasing function satisfying $\bar{\alpha}(0)=0$. For instance, we can always consider redefinitions of the form
\begin{equation}\label{eq:lredef}
\ell\longrightarrow \bar{\ell}=\ell^p M^{1-p}= M (\ell/M)^p,\qquad p>0.    
\end{equation}
Under such a redefinition, we have
\begin{equation}
    \alpha\longrightarrow \bar{\alpha}=\frac{\bar{\ell}}{M}=\left(\frac{\ell}{M}\right)^{p}=\alpha^{p}.
\end{equation}
Note that both $\alpha$ and $\bar{\alpha}$ take values in $[0,+\infty)$ due to $p>0$. Then, we see that condition \ref{cond3}, which assumes the existence of a redefinition of the form~\eqref{eq:lredef} for which
\begin{equation}
f(r,M,\ell)=f(r,\bar{\ell}),
\end{equation}
is equivalent to
\begin{equation}\label{eq:cond3m}
f(r,M,\ell)=f(r,\ell^p M^{1-p})= f( r/[\ell^p M^{1-p}]),\qquad p>0,
\end{equation}
which hence can be used as an alternative form for the same condition.

\subsection{Non-singular black holes} \label{sec:rbh}

Non-singular black holes are characterized by the condition that the time-independent version of Eq.~\eqref{eq:horloc},
\begin{equation}\label{eq:horlocnov}
 2m(r_{h_i})=r_{h_i},   
\end{equation}
has at least one root to accommodate the existence of an outer horizon as defined by Hayward~\cite{Hayward1994,Hayward:2005gi}. The existence of at least one root, together with regularity at $r=0$, requires the existence of an even number of roots (counting multiplicities). This is a corollary of Eqs.~\eqref{eq:metreg} and~\eqref{eq:metaflat}. 

Without loss of generality, we can focus our attention on the simplest case in which there are two roots. If these roots are simple, these mark the location of outer and inner horizons~\cite{Hayward1994,Hayward:2005gi}. If these roots have the same value, being therefore a double root, the corresponding horizon is extremal in the sense that the associated surface gravity vanishes. It is also possible to have intermediate situations in which both horizons have distinct positions, but one of them is extremal~\cite{Carballo-Rubio:2022kad}.

The two roots, $r_+(\alpha)$ and $r_-(\alpha)$, are functions of the dimensionless parameter $\alpha=\ell/M$ introduced above. We are using the naming convention that $r_-(\alpha)\leq r_+(\alpha)$, so that $r_+(\alpha)$ corresponds to the outer horizon and $r_-(\alpha)$ to the inner horizon.

These roots recover their values for the Schwarzschild geometry in the limit $\alpha\rightarrow0$:
\begin{equation}   \lim_{\alpha\rightarrow 0}r_-(\alpha)=0,\qquad\qquad \lim_{\alpha\rightarrow 0}r_+(\alpha)=r_{\rm S}=2M.
\end{equation}
The generic presence of an inner horizon is particularly relevant as it casts concerns regarding the viability of such geometries. In fact, it has been shown that regular black holes suffer from mass inflation instability \cite{Carballo-Rubio:2018pmi,Carballo-Rubio:2021bpr,DiFilippo:2022qkl}. Due to the large blueshift at the inner horizon, a small perturbation  has a large backreaction on the geometry, possibly destabilizing it. This property is extremely general as it can be derived with purely geometrical considerations. However, it is only possible to prove the presence of a linear instability, while the non-linear analysis would require the knowledge of the dynamics of the theory leading to regular black holes. The result of the non-linear analysis can be either a migration to a horizonless configuration, or a regular black hole with zero surface gravity at the inner horizon, which would not suffer from mass inflation instability \cite{Carballo-Rubio:2022kad}. Regardless of this limitation, the mass inflation instability teaches us that there is still a lot of work to do before we can claim that regular black holes provide a complete alternative representation of black holes.

\subsection{Horizonless stars}\label{sec:horizonless}

In contrast to regular black holes, horizonless stars are characterized by the condition that Eq.~\eqref{eq:horlocnov} does not have any roots. This guarantees the lack of horizons of any kind in the geometry. The metric we are considering is not generally a vacuum solution of the Einstein field equations, even if it asymptotically approaches the Schwarzschild geometry. This allows us to define an effective radius of the source for which the deviations from the Schwarzschild geometry become $\mathcal{O}(1)$~\cite{Carballo-Rubio:2018jzw}. This effective radius $R$ can be parameterized as $R=r_{\rm s}(1+\Delta)$, where $r_{\rm s}$ is the Schwarzchild radius and $\Delta\geq 0$. Thus for the ``compactness" of the star we have $\chi = 2M/R = 1/(1+\Delta)$. The smaller the value of $\Delta$, the more compact the geometry is considered, as the effective size of the source becomes smaller. Aside from that, horizonless stars can still have tails for $r>R$, these tails being defined as modifications of the geometry that decay with radial distance, typically following a polynomial law~\cite{Carballo-Rubio:2018jzw}.

These geometries are horizonless, meaning that $g_{vv}(r)$ never vanishes, and therefore there are no hypersurfaces of infinite redshift. However, $g_{vv}(r)$ can take arbitrarily small values depending on the value of $\Delta$, which means that the maximum redshift attained, being the inverse of the former dimensionless quantity, can be arbitrarily large. To obtain this maximum redshift, we just need to evaluate the minimum of the function $1/g_{vv}(r)$. We will calculate this for an specific example later, showing in particular that, for $\Delta\ll1$, the maximum redshift scales as $1/\Delta$.

There is an interesting parallelism between horizonless stars and regular black holes. While horizonless stars do not have an inner horizon, and therefore do not display the associated instability discussed in the previous section, it has been shown that horizonless stars must have a even number of light rings, one of which is stable \cite{Cunha:2020azh}. The presence of a stable light ring is often associated with an unstable configuration, as massless perturbations would pile up there and eventually their backreaction on the geometry would not be negligible \cite{Cunha:2022gde}. Contrary to the regular black hole case, this instability is non-linear and the analysis has to be carried out case by case in a model-dependent way. Recently, the presence of such instability has been confirmed for some specific models of boson and Proca stars, in which the presence of the stable light ring destabilizes the geometry in a short timescale leading to either a non compact configuration or a collapse into a black hole \cite{Cunha:2022gde}. More analyses in different models of compact objects are required to understand if this instability is a generic feature or if there are stable compact horizonless models. In particular, one aspect that needs to be analyzed is that the standard argument requires that matter follow null geodesics and can pile up at the light ring. This assumption is reasonable if there are no interactions between matter and the central object. However, it is straightforward to realize that the stable light ring is always contained within the non-vacuum region of the geometry, which means that any interactions between the accreting matter and the matter supporting the compact object (possibly mediated by the gravitational interaction) can take over as the dominant process and render the instability meaningless.

Due to their quite different spacetime structures, horizonless stars and non-singular black holes are typically considered as unrelated kinds of geometries. However, as we discuss below, these two classes of geometries are in fact complementary subsets of the families of geometries we are considering. In fact, each of the families considered above can be spit into these two subsets (regular black holes and horizonless stars), with a marginal case (extremal regular black hole) as the boundary between the two subsets.

\subsection{Horizonless stars associated with non-singular black holes}

In the discussion above, we have discussed how the structure of roots of $g_{vv}(r)$ [equivalently, Eq.~\eqref{eq:horlocnov}] determines whether the spacetime being considered is either a regular black hole or a horizonless object. This is related to the values that the parameter $\alpha$ takes. We discuss this explicitly below. We keep working with static geometries; general dynamical situations will be analyzed in Sec.~\ref{sec:dyn}.

The spacetimes we are studying are characterized by a function $m(r)$ taking the form in Eq.~\eqref{eq:cond3}. We have already discussed that it is useful to introduce dimensionless variables $\bar{\alpha} =\bar{\ell} /M\in [0,+\infty)$ and $x =r_{h_i} /\bar{\ell}\in [0,+\infty)$, as well as the function $h(x) = x/2f(x)$, in terms of which Eq.~\eqref{eq:horloc1} takes the form
\begin{equation}
\label{E:roots}
\frac{1}{\bar{\alpha}} = h(x).
\end{equation}
Assuming that the Misner--Sharp--Hernandez mass $m(r)$ is non-negative, the function $f(x)$ must also be non-negative in its domain of definition $x\in[0,+\infty)$. Then, $h(x)$ is bounded from below, and we can define
\begin{equation}
h_\star = \inf_{x\geq 0} h(x) = \inf_{x\geq 0} \frac{x}{2f(x)} \geq 0.
\end{equation}
\begin{itemize}
\item The equation (\ref{E:roots}) has  zero solutions if $\bar{\alpha}> 1/ h_\star$. The corresponding spacetime describes a horizonless object.
\item The equation (\ref{E:roots}) has one or more solutions if $\bar{\alpha}= 1/ h_\star$. The corresponding spacetime describes an extremal regular black hole.
\item The equation (\ref{E:roots})  has one or more solutions if $\bar{\alpha}< 1/ h_\star$. The corresponding spacetime describes a regular black hole.
\end{itemize}

Hence, we see that there always exists a critical value $\bar{\alpha}=\bar{\alpha}_\star=1/h_\star$ that divides the family in two different subsets. Geometries with $\bar{\alpha}\in(0,\bar{\alpha}_\star)$ describe regular black holes, while geometries with $\bar{\alpha}\in(\bar{\alpha}_\star,+\infty)$ describe horizonless stars. There is a limiting case $\bar{\alpha}=\bar{\alpha}_\star$ describing an extremal regular black hole. 

\begin{figure}
    \centering
    \includegraphics[width=0.75\columnwidth]{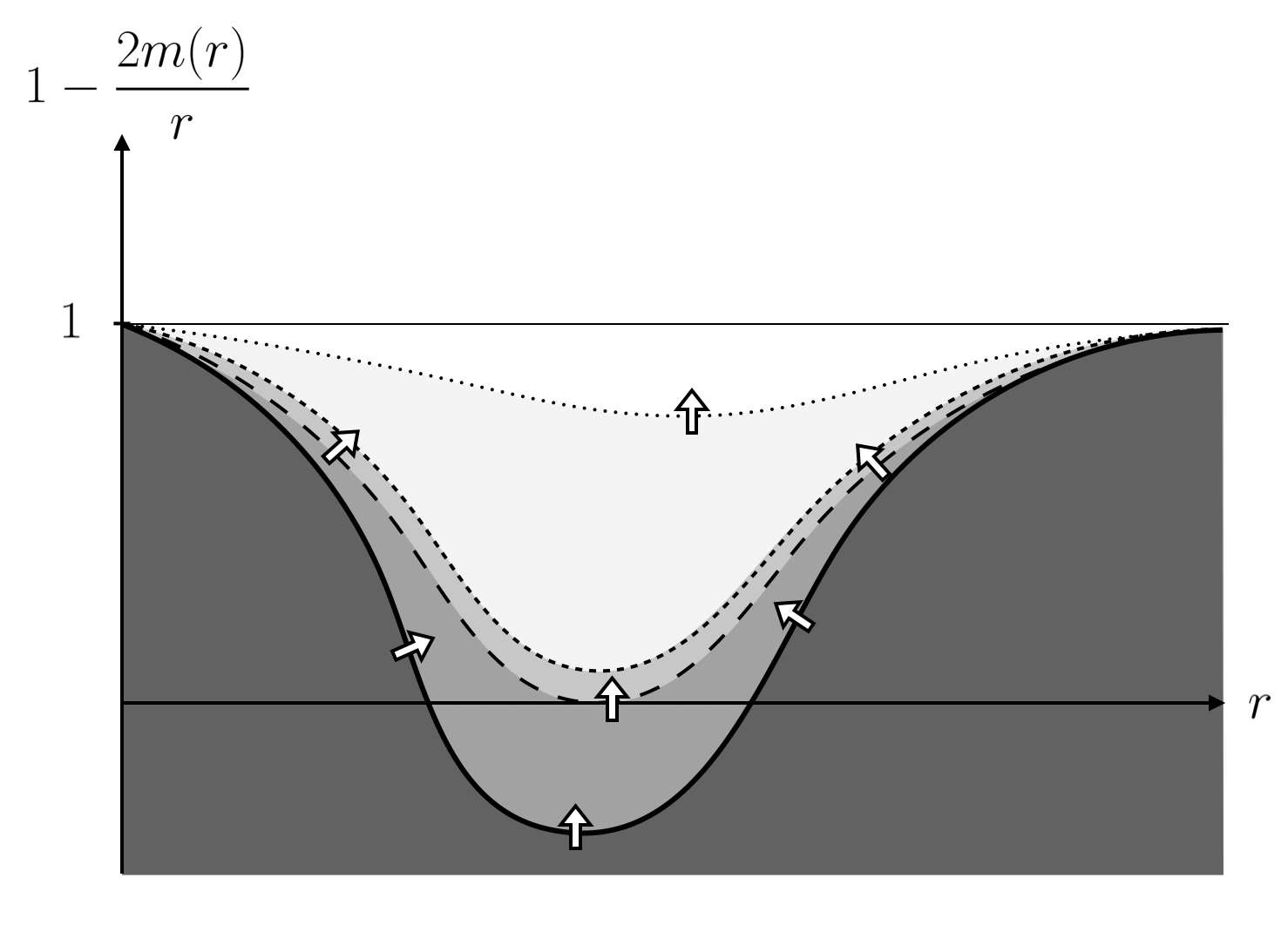}
    \caption{This figure illustrates how non-singular black hole geometries with $\bar{\alpha}\in[0,\bar{\alpha}_{\star}]$ are deformed into horizonless configurations for $\bar{\alpha}\in(\bar{\alpha}_{\star},+\infty]$. These horizonless configurations are ultracompact for $\bar{\alpha}-\alpha_{\star}\ll 1$ but become diluted (that is, the effective radius increases) for increasing values of $\alpha$. In fact, for $\Phi=0$ the $\bar{\alpha}\rightarrow\infty$ limit is the Minkowski spacetime. The solid line indicates a non-singular black hole ($\bar{\alpha}<\bar{\alpha}_{\star}$), the sparsely dashed line an extremal non-singular black hole ($\bar{\alpha}=\bar{\alpha}_{\star}$), the thinly dashed line an ultracompact horizonless object ($\bar{\alpha}> \bar{\alpha}_{\star}$, $\bar{\alpha}-\bar{\alpha}_{\star}\ll 1$), and the dotted line a low-density horizonless configuration ($\bar{\alpha}> \bar{\alpha}_{\star}$, $\bar{\alpha}-\bar{\alpha}_{\star}\gg 1$). The white arrows indicate how the functional profile of $m(r)$ changes when $\bar{\alpha}$ increases. \label{fig:nsbh_to_eco}}
\end{figure}

The compactness of horizonless stars is a function of $\bar{\alpha}\in[\bar{\alpha}_\star,+\infty)$, so that this compactness is higher for $\bar{\alpha}-\bar{\alpha}_\star\ll 1$ while, in the limit $\bar{\alpha}\rightarrow\infty$ these objects become diluted and, in fact, Minkowski spacetime is recovered (see Fig.~\ref{fig:nsbh_to_eco}).

\section{Explicit example: horizonless stars of Hayward type}\label{eq:haywex}

In this section, we consider a specific example in order to illustrate our general discussion above. Let us consider the static Hayward metric corresponding to the choices~\cite{Hayward:2005gi}:
\begin{equation}\label{eq:Hayward}
m(r)=\frac{Mr^3}{r^3+2\ell^2 M},\qquad \phi(r)=0,    
\end{equation}
thus defining a specific two-parameter family of geometries $\{g_{ab}(M,\ell)\}_{M\in\mathbb{R}^+,\ell\in\mathbb{R}^+}$. The parameter $M$ is the Arnowitt--Deser--Misner (ADM) mass, while $\ell$ is a parameter with dimensions of length that controls the value of curvature invariants around the center of spherical symmetry~\cite{Hayward:2005gi} as well as the effective stress-energy tensor sourcing the geometry. This metric satisfies Eq.~\eqref{eq:cond3m} for the specific value $p=2/3$. That is, herein $\bar\ell = \sqrt[3]{M \ell^2} = M (\ell/M)^{2/3}$.

This effective stress energy tensor takes the form associated with an anisotropic perfect fluid, with:
\begin{equation}\label{eq:effsour}
\rho(r) =\frac{3\ell^2}{2\pi}\left(\frac{m(r)}{r^3}\right)^2= -p_r(r),\qquad p_t(r) = \frac{3\ell^2}{\pi}\frac{r^3-\ell^2 M}{r^3+2\ell^2 M}\left(\frac{m(r)}{r^3}\right)^2=\frac{2r^3-2\ell^2 M}{r^3+2\ell^2 M}\rho(r).
\end{equation}

We note that
\begin{equation}
    w_r \equiv {p_r\over\rho} \equiv -1; \qquad 
    w_t \equiv {p_t\over \rho} = {2(r^3-\ell^2M)\over r^3 +2\ell^2 M} \in [-1,2].
\end{equation}
Thence 
\begin{equation}
    w_t-w_r = {3r^3\over r^3 + 2\ell^2 M } \geq 0,
\end{equation}
which makes it clear that the pressures are anisotropic everywhere away from the centre at $r=0$. 

\begin{figure}
    \centering
    \includegraphics[width=0.7\columnwidth]{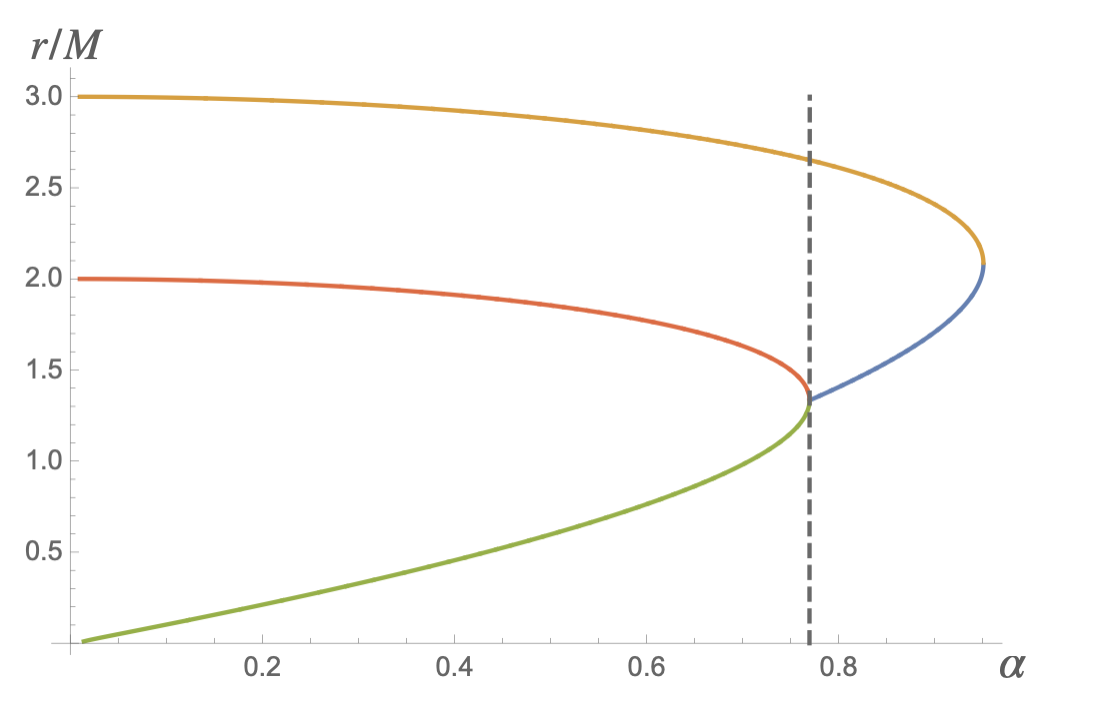}
    \caption{Location of the inner horizon (green), outer horizon (red), inner light ring (blue) and outer light ring (yellow) for different values of $\alpha$. The dashed line corresponds to $\alpha=\alpha_\star$.
    \label{fig:HLR}}
\end{figure}

As discussed above on general grounds, the relative values of $\ell$ and $M$ or, more precisely, the dimensionless quotient $\alpha=\ell/M$, determine the structure of horizons of these geometries. Studying the roots of Eq.~\eqref{eq:horloc} shows that there are two single roots if $\alpha< 4/3\sqrt{3}$, a double root  if $\alpha= \alpha_\star=4/3\sqrt{3}$ (at $r_\star=4M/3= \sqrt{3}\ell$), and no roots if $\alpha> 4/3\sqrt{3}$. This is precisely the situation that we have described generically before, for a particular value of $\alpha_\star$ that arises from the specific choice of functions in Eq.~\eqref{eq:Hayward}. The presence of light rings also depends on $\alpha$. For small values of $\alpha$ there is only one light ring. For $\alpha\geq\alpha_\star$, a second light ring is formed. In particular, for $\alpha=\alpha_\star$ the location of the inner light ring coincides with the degenerate horizon. The situation is summarised in Fig.~\ref{fig:HLR}. The presence of an odd numbers of light rings for a regular black hole is not peculiar to the Hayward geometry. Appendix \ref{sec:LR}, shows that this results is true for any spherically symmetric static regular black hole. 

The original motivation of Hayward in~\cite{Hayward:2005gi} was constructing an effective model for the formation and evaporation of regular black holes, taking $\ell$ to be the Planck length. Under this assumption, a macroscopic value of the mass $M$ (e.g.~one solar mass) would imply the separation of scales $\ell\ll M$. Hayward assumed that the leading dynamical effects in regular black holes could be described by promoting $M$ to a time-dependent function, while keeping the value of $\ell$ fixed. This is an assumption regarding the time evolution of this metric, that we will put under scrutiny in Sec.~\ref{sec:dyn}.

\begin{figure}
    \centering
    \includegraphics[width=0.45\columnwidth]{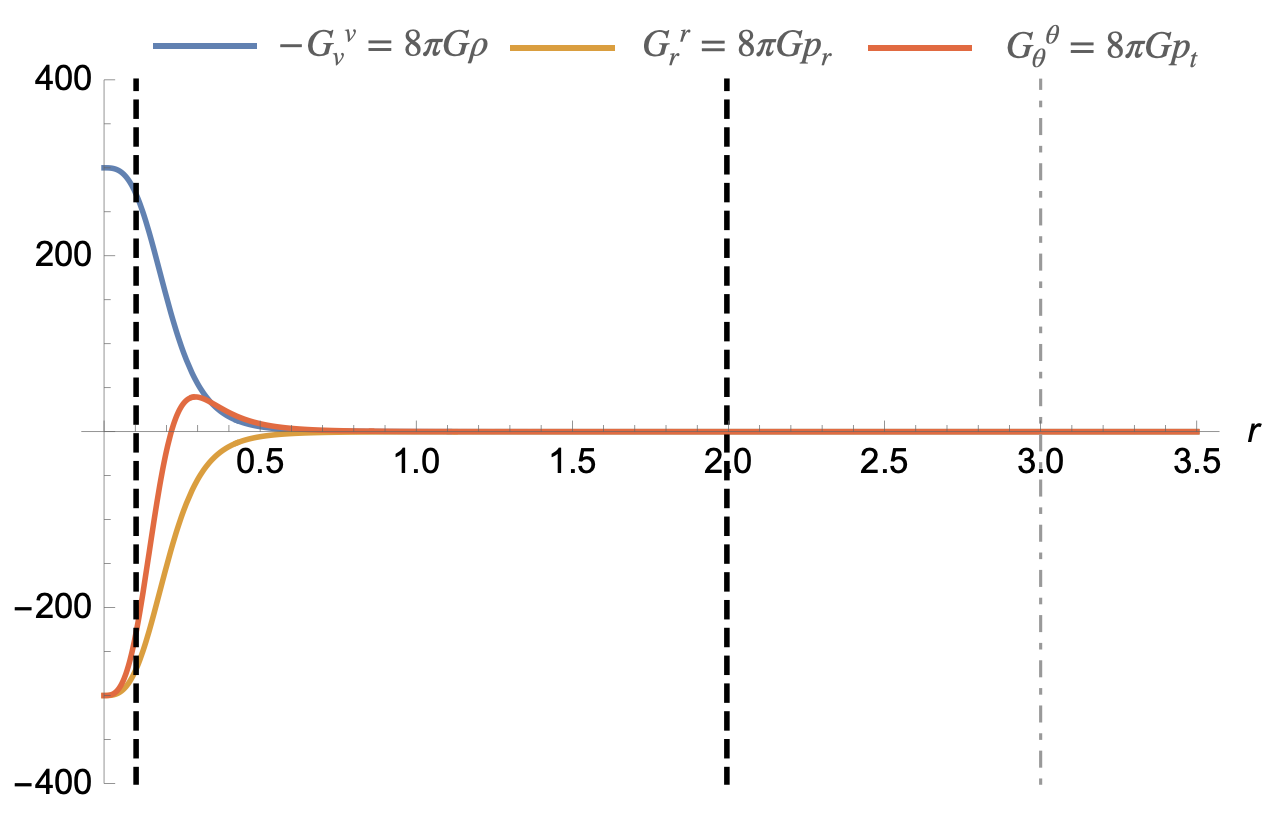}\qquad
    \includegraphics[width=0.45\columnwidth]{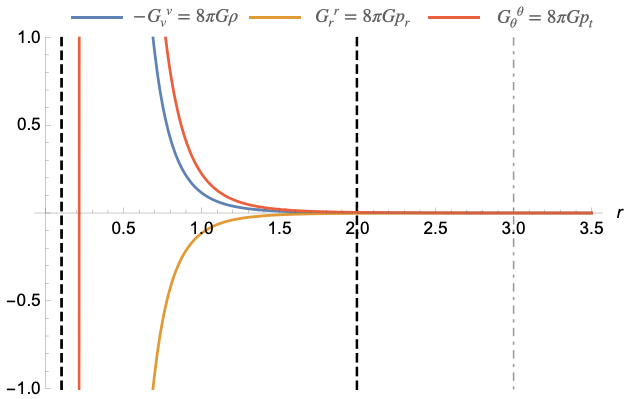}
    \caption{Value of the relevant components of the Einstein tensor for $M=1$ and $\alpha=0.1$. The dashed black lines mark the inner and outer horizons, while the grey dot-dashed line correspond to the light ring. The left figure shows that within the inner horizon the effective energy density and pressures go to constant values for small $r$ signaling the presence of a de Sitter core. The right figure plots the same quantities on a different scale to show that deviations from Schwarzschild geometry are strongly suppressed at the outer horizon. 
    \label{fig:einst_small_l}}
\end{figure}

\begin{figure}
    \centering
    \includegraphics[width=0.7\columnwidth]{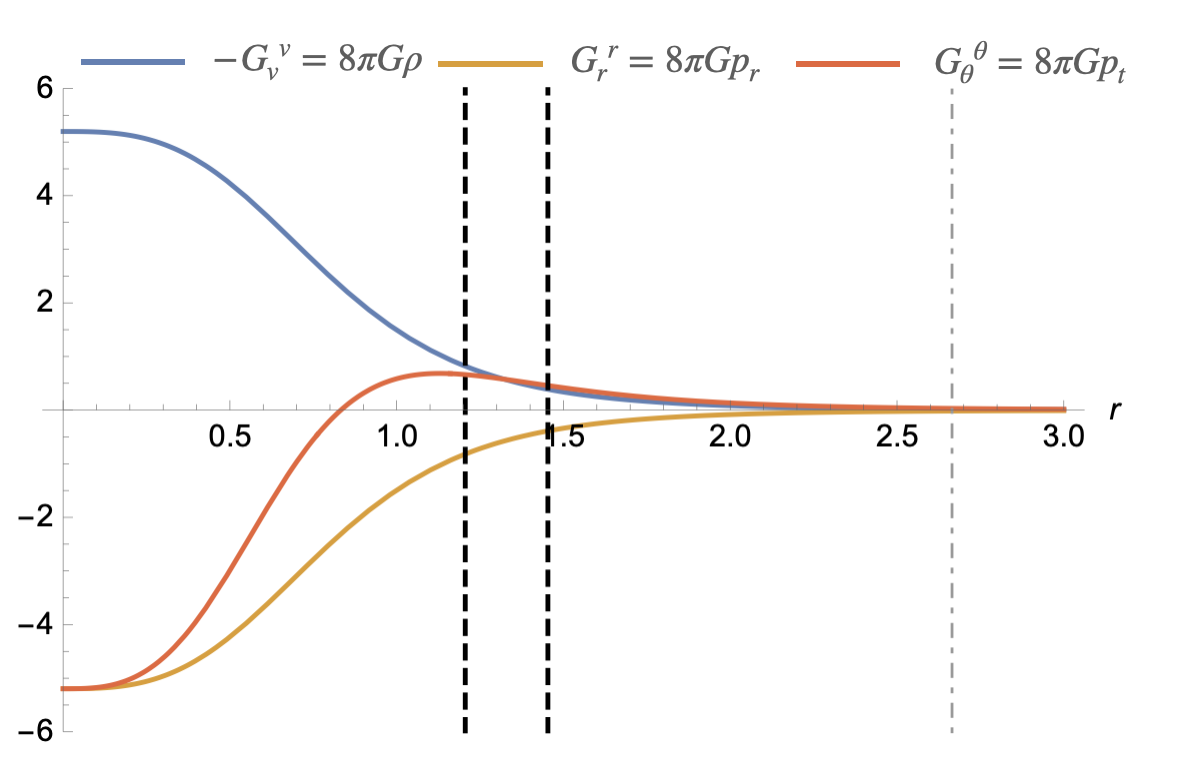}
    \caption{Value of the relevant components of the Einstein tensor for $M=1$ and $\alpha=\alpha_\star-0.01$. The dashed lines mark the inner and outer horizon, while the grey dot-dashed line correspond to the light ring. The effective energy density and pressures go to constant values for small $r$ signaling the presence of a de Sitter core. Deviations from  Schwarzschild geometry decay at large distances, but they are still non-negligible at the gravitational radius.
    \label{fig:einst_large_l}}
\end{figure}

\begin{figure}
    \centering
    \includegraphics[width=0.7\columnwidth]{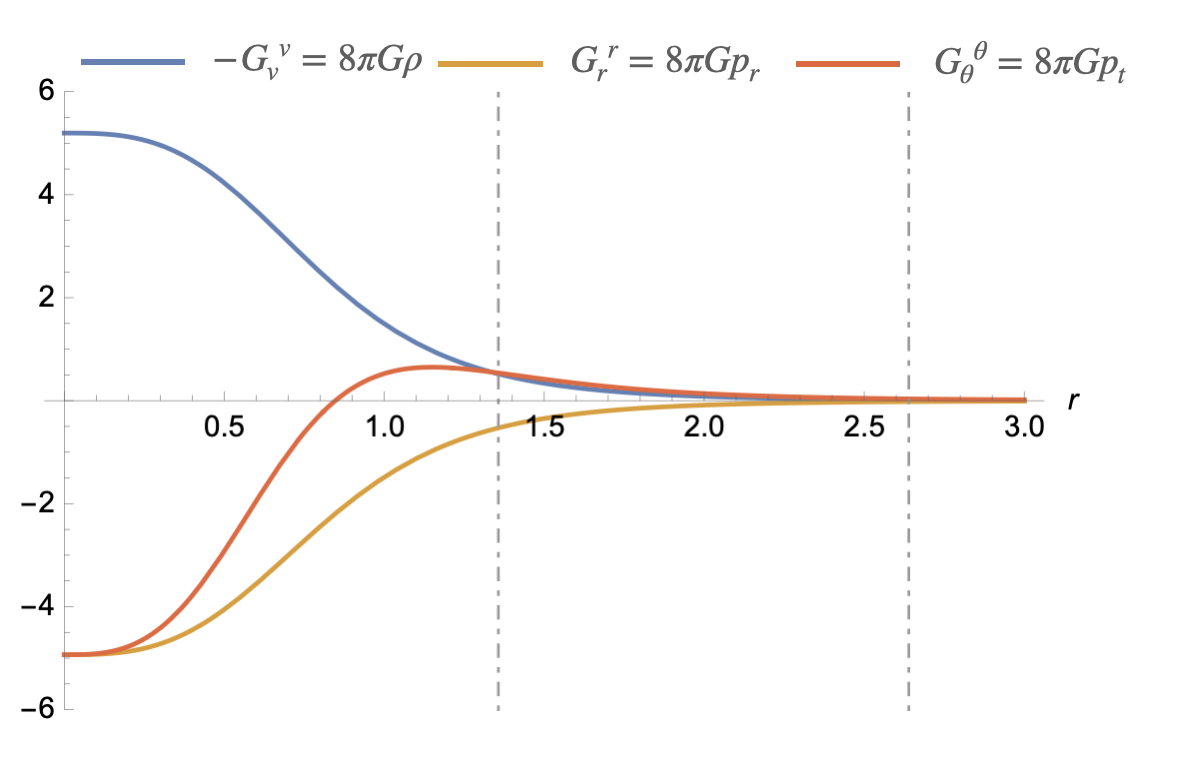}
    \caption{Value of the relevant components of the Einstein tensor for $M=1$ and $\alpha=\alpha_\star+0.01$. The geometry does not possess horizons, beside the outer light ring, a second light ring is present. The two light rings are marked by the grey dot-dashed lines. The inner light ring is located in a region where the effective energy and pressures are non-negligible.
    \label{fig:largelco}}
\end{figure}

Let us consider two different situations, one in which this separation of scales takes place, and another in which both scales are comparable. As we have discussed above, it is more adequate to use the parameter $\alpha$ to describe these situations. In terms of this dimensionless parameter, we can focus on two different situations of special interest:
\begin{itemize}
    \item $\alpha\ll 4/3\sqrt{3}$: as illustrated in Fig.~\ref{fig:einst_small_l}, the corresponding geometries describe a structure with a de Sitter core within the inner horizon, surrounded by a transition region from the de Sitter core to a Schwarzschild geometry. Around the gravitational radius, deviations from the Schwarzschild geometry are already strongly suppressed by a factor $\alpha^4$, and decay polynomially with the radius.

 \item $\alpha\sim	 4/3\sqrt{3}$:  as illustrated in Fig.~\ref{fig:einst_large_l} and Fig.~\ref{fig:largelco} the de Sitter core is still within in the interior region, but deviations from the Schwarzschild geometry become $\mathcal{O}(1)$ at the gravitational radius. The qualitative features of this effective source are reminiscent of gravastars~\cite{Chapline:2000en,Mazur:2001fv,Mazur:2004fk,Visser:2003ge}, more specifically with models that avoid the introduction of infinitesimally thin shells so that the pressure is continuous~\cite{Cattoen:2005he}. These similarities include the equation of state for the radial pressure being $p_r=-\rho$, with a positive energy density, and with $p_r(0)=p_t(0)=\rho(0)$ at the center of spherical symmetry, with the pressure being anisotropic elsewhere, which is necessary for gravastar-like configurations~\cite{Cattoen:2005he}. It is important to distinguish the cases $\alpha\lesssim	 4/3\sqrt{3}$ and $\alpha\gtrsim 4/3\sqrt{3}$. In the first case, the geometry has both outer and inner horizons and a light ring (see Fig.~\ref{fig:einst_large_l}).  In the second case, the geometry is now horizonless and possess a second light ring which is located in a region where the effective stress energy tensor is non-negligible (see Fig.~\ref{fig:largelco}).  
 \end{itemize}
When $\alpha$ takes values in between the two situations above, the corresponding geometries describe intermediate situations in which the de Sitter core becomes larger as $\alpha$ increases. These intermediate situations play an essential role in the dynamical considerations discussed in the next section. For values of $\alpha$ greater than the ones considered in the second case, the corresponding horizonless stars become more dilute progressively until becoming indistinguishable from Minkowski spacetime in the $\alpha\rightarrow\infty$ limit. We will not discuss further these situations as we will mostly focus on ultracompact horizonless stars, but these less compact configurations could still play an interesting role in phenomenological studies.

\section{Dynamical considerations}\label{sec:dyn}

In the previous section we have discussed specific static examples. The fact that the family of static geometries considered contains both non-singular black holes and horizonless stars suggests that it may be possible to find dynamical geometries in which non-singular black holes evolve towards horizonless stars. Here we aim at understanding whether semiclassical physics contain the seeds for this evolution, by analyzing the effective stress-energy tensor associated with these interpolating dynamical geometries.

Let us consider for simplicity the metric~\eqref{eq:spslinel} with $\Phi(v,r)=0$,
\begin{equation}\label{eq:spslinelphi0}
        \text{d}s^2=-\left(1-\frac{2m(v,r)}{r}\right)\text{d}v^2+2\text{d}v\text{d}r+r^2\text{d}\Omega^2.
\end{equation}
This is sometimes called generalized Vaidya spacetime. The time independence is introduced by promoting $M$ and $\ell$ to functions of $v$, so that the time dependence on $m(v,r)$ is implicit.

It is certainly possible to prescribe functions $M(v)$ and $\ell(v)$ that lead to a transition between regular black holes and horizonless stars. This require that $\ell$ flows from an initial value $\ell_0$ to a value greater than $\alpha_{\rm crit} M(v)$, where the specific value of $\alpha_{\rm crit}$ depends on the specific model (for instance, $\alpha_{\rm crit}=4/3\sqrt{3}$ for the Hayward model). The goal of this section is to understand here is whether such evolution can be justified within the framework of semiclassical physics.

The non-zero components of the effective stress energy tensor associated with the metric in Eq.~\eqref{eq:spslinelphi0} are given by

\begin{equation}
    T_v\,^v=T_r\,^r=-\frac{1}{4\pi r^2}\frac{\partial m(v,r)}{\partial r},\qquad 
    T_\theta\,^\theta=T_\phi\,^\phi=-\frac{1} {8\pi r}\frac{\partial^2 m(v,r)}{\partial r^2},\qquad 
    T_v\,^r=\frac{1}{4\pi r^2}\frac{\partial m(v,r)}{\partial v}.
\end{equation}
Ordering the coordinates as $(v,r,\theta,\phi)$ we can write it in the matrix notation as

\begin{equation}
  T_a{}^b = -{1\over 8\pi r^2} \left[ \begin{array}{cc|cc}
2m' & -2\dot m &0&0\\
0 & 2m' &0 &0\\
\hline
0 &0& r m'' &0\\
0 &0& 0 & r m''
\end{array}
\right].  
\end{equation}
This is of non-trivial Jordan normal form, and specifically, guarantees that the stress--energy tensor is of Hawking--Ellis type II~\cite{essential-core}. 
Indeed one has $T_a{}^b = T^{I\,b}_a + \dot m k_a k^b/4\pi r^2$, where $k~a$ is a suitable outward-pointing null vector is $k^a=(0,1,0,0)$, with corresponding co-vector $k_a = (1,0,0,0)$, and $T^{I\,b}_a$ is type I (thus diagonal).

Let us analyze two different situations:
\begin{itemize}

\item If there is no explicit time dependence, $m=m(r)$ and the only non-vanishing components are $T_v\,^v$, $T_r\,^r$, $T_\theta\,^\theta$, $T_\phi\,^\phi$. These components indicate the stresses that are necessary to maintain the geometric structure, for instance of a static regular black hole.

\item When switching on time dependence, the components of the stress-energy tensor split naturally into two kinds of components. The components $T_v\,^v$, $T_r\,^r$, $T_\theta\,^\theta$, $T_\phi\,^\phi$ do not get any additional contributions and remain the same from both functional and conceptual perspectives, and describe how the stresses necessary to maintain the time-dependent structure change in time. Aside from this, there is an additional flux $T_v\,^r$, that can be written as
\begin{equation}\label{eq:flux}
T_v\,^r=\frac{1}{4 \pi r^2}
\left(\frac{\partial m(v,r)}{\partial M(v)}\; \frac{\partial M(v)}{\partial v}
+\frac{\partial m(v,r)}{\partial \ell(v)}\;
\frac{\partial \ell(v)}{\partial v}
\right).
\end{equation}

\end{itemize}

\subsection{Time dependence in the ADM mass}

Let us now compare the expressions of the stress energy tensor obtained when we promote only one of the two quantities  $M$ and $\ell$ to be a function of time. From Eq.~\eqref{eq:flux}, we learn that the radial profile of the flux does not depend on the actual time dependence we show for these parameters, but rather on the quantities $\partial m /\partial M$ and $\partial m/\partial \ell$. Hence, an analysis of these quantities can contain valuable information about the radial behavior of the effective source in these geometries.

Let us starting by promoting $M$ to a function $M(v)$. Given that the quasilocal mass function $m(v,r)$ has to approach the value of the asymptotic mass $M(v)$ for large values of the radius, we have
\begin{equation}
   \lim_{r\rightarrow\infty} \frac{\partial m(v,r)}{\partial M(v)}=1\,.
\end{equation}
Therefore, at large distances we have
\begin{equation}
    T_v\,^r=\frac{1}{4 \pi r^2} \dot{M}+\mathcal{O}\left(r^{-3}\right)\,,
\end{equation}
where $\dot{M}=\partial M/\partial v $.

This behaves as a flux of radiation for $r\rightarrow\infty$ (decays as $1/r^2$). The flux is positive if $\dot{M}(v)>0$ and negative if $\dot{M}(v)<0$.
On the other hand, the value of $T_v\,^r$ for small values of $r$ depends on the choice of the geometry. For concreteness, let us consider the Hayward geometry \eqref{eq:Hayward}.
We have then 
\begin{equation}
    T_v\,^r=\frac{1}{4\pi} \frac{r^4 \dot{M}}{(r^3+2\ell^2 M)^2}
\end{equation}
As discussed before, this behaves as a flux of radiation for $r\rightarrow\infty$. 
The value of $r=r_{\star}$ for which this flux is maximal is determined by the relation
\begin{equation}
4r_{\star}^3(r_{\star}^3+2\ell^2 M)-6r_{\star}^6=0,    
\end{equation}
namely
\begin{equation}\label{eq:mfluxmax}
r_{\star}=(\ell^2M)^{1/3}. 
\end{equation}
The radial profile associated with this flux is displayed in Fig.~\ref{fig:T01}.

\subsection{Time dependence in the core size}

Let us now consider the case in which we only promote $\ell$ to a function $\ell(v)$. As before, the asymptotic behavior is fixed by the requirement that $m(v,r)$ at large distances approaches $M$ and it is therefore independent of $\ell$
\begin{equation}
    \lim_{r\rightarrow\infty}\frac{\partial m}{\partial\ell}=0\,.
\end{equation}
Therefore, $T_v\,^r$ decays more rapidly than a flux of radiation. Considering again the Hayward geometry, we have
\begin{equation}
    T_v\,^r=-\frac{1}{\pi} \frac{rM^2\ell\dot{\ell}}{(r^3+2\ell^2 M)^2}\,,
\end{equation}
where $\dot{\ell}=\partial\ell/\partial v$.
Which, as expected, decays more rapidly than a flux of radiation for $r\rightarrow\infty$ (namely as $1/r^5$). Note also that there is a relative minus sign with respect to the case discussed previously. The flux is positive if $\dot{\ell}(v)<0$ and negative if $\dot{\ell}(v)>0$.

\begin{figure}
    \centering
    \includegraphics[scale=.65]{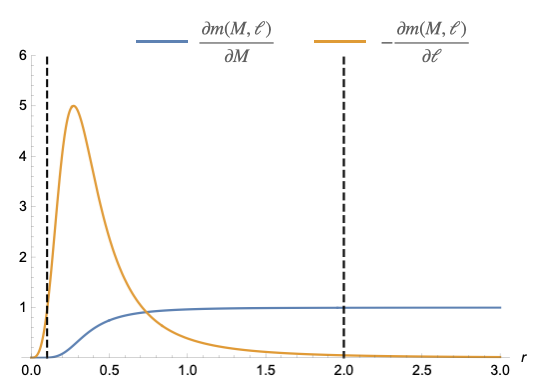}
    \caption{Values of $\partial m/\partial M$ and $\partial m/\partial \ell$ which are proportional to the flux required to promote either $M$ or $\ell$ to a function of time. We considered the Hayward geometry evaluated at the values $M=1$ and $\ell=0.1$. The dotted lines correspond to the inner and outer horizon. The flux associated to the time dependence of the ADM mass $M$ monotonically increases from $0$ to a constant value, while the flux associated to the time dependence of the regularization parameter $\ell$ goes to zero both at the center and asymptotically and it reaches a maximum value slightly outside the inner horizon. }
    \label{fig:T01}
\end{figure}

The value of $r=r_{\star}$ for which this flux is maximal is determined by the relation
\begin{equation}
r_{\star}^3+2\ell^2 M-6r_{\star}^3=0,    
\end{equation}
namely
\begin{equation}
r_{\star}=\sqrt[3]{\ell^2M/5}. 
\end{equation}
This maximum is at a different position than the maximum in Eq.\eqref{eq:mfluxmax}. The differences between the radial behavior of both fluxes is illustrated more clearly in Fig.~\ref{fig:T01}.
%
\subsection{Summary}

In the previous two sections, we have discussed the effective source associated with two different kinds of situations:
\begin{itemize}
    \item Promoting $M$ to a function $M(v)$: this is a standard choice in the literature (e.g.~\cite{Hayward:2005gi}) to describe the evaporation of black holes. It is reasonable to assume that a time dependence on this sort is present, and the corresponding effective source is consistent with semiclassical physics from around the outer horizon to the asymptotic region.
    \item Promoting $\ell$ to a function $\ell(v)$: this possibility has not previously been considered, to the best of our knowledge, and we have found that the effective source associated with it takes qualitatively the same form as the one in the item above, but just with a different radial profile so that this effective source is localized around the inner horizon.
\end{itemize}

Both points above imply that assuming that $\ell$ remains constant is a strong assumption, given our lack of knowledge of semiclassical backreaction in the interior of regular black holes. In other words, there is no proof that just promoting $M$ to a function $M(v)$ provides a good effective description of semiclassical physics. It is necessary to analyze carefully the semiclassical backreaction around the inner horizon to determine whether or not the standard picture of evaporation involves also a time dependence of the form $\ell(v)$, and therefore a displacement of the inner horizon. The only studies of this issue that have been performed so far show results that are indeed consistent with an evolution of this length scale \cite{Barcelo:2020mjw,Barcelo:2022gii}, so that the inner horizon is displaced outwards. This is also consistent with what one would expect on the basis of the instability of inner horizons~\cite{Carballo-Rubio:2018pmi,Carballo-Rubio:2021bpr}\footnote{An alternative possibility is that the surface gravity evolves dynamical towards a vanishing value, resulting in the geometry proposed in~\cite{Carballo-Rubio:2022kad}. We leave a detailed discussion of this possibility for future work.}. As we have discussed in this paper, this could be the first steps in the dynamical evolution of regular black holes towards horizonless ultracompact stars. Hence, we think that a careful analysis of semiclassical physics around inner horizons, and not only outer horizons, is of fundamental importance in order to understand the structure of black holes.

\section{Conclusions}

In this paper we have shown that each family of regular black holes has a complementary family of horizonless stars, a subset of which are ultracompact. This connection has never been identified in the literature to the best of our knowledge and, in fact, these different objects were considered to describe alternative end states of black holes once the new physics necessary to get rid of singularities is included in the picture.

We also stress that our analysis clarifies how the inner/outer horizon structure intrinsic to the regular black hole geometries is deformed continuously into an inner/outer light ring structure for the exterior of the ultracompact stars obtained as the complementary geometries to the regular black hole geometries. It is also interesting to note that both inner horizons and light rings pose similar viability issues for the corresponding models, which strongly suggests that some form of dynamics is needed to characterize at least the early times of these objects.

Indeed, our work provides further evidence that these different geometries may in fact represent different phases of black holes during their lifetime. We have analyzed dynamical situations in which regular black holes evolve towards horizonless stars, and show that it is plausible that the effective stress-energy tensor that results in the first stages of evolution is compatible with semiclassical physics. Hence, gravitational collapse may lead first to a regular black hole, that could evolve dynamically towards an ultracompact horizonless star. The details of such a transition need to be fleshed out, but we think the evidence for the plausibility of this scenario is strong enough to call for future studies of this issue.

\bigskip
\bigskip
\hrule\hrule\hrule


\appendix
\section{Horizonless stars of the Bardeen type}

\noindent
While in this article we have concentrated on the Hayward black hole it is worth asking whether looking at the Bardeen~\cite{Bardeen1968} regular black hole teach us anything fundamentally new.
Indeed, the key differences for the Bardeen regular black hole~\cite{Bardeen1968} are that the Misner--Sharp quasi-local mass is now replaced by
\begin{equation}
    m(r) = {M r^3\over (r^2+\ell^2)^{3/2}},
\end{equation}
implying that for the Bardeen regular black hole one can conveniently set $\bar\ell=\ell$.

The density and principal pressures become
\begin{equation}
    \rho = - p_r = {3M\ell^2\over4\pi (r^2+\ell^2)^{5/2} };
    \qquad\qquad
    p_t = {3M\ell^2(3 r^2-2 \ell^2) \over 8\pi (r^2+\ell^2)^{4/2}}.
\end{equation}
Then for the $w$ parameter:
\begin{equation}
    w_r=-1; \qquad w_t = {3r^2-2\ell^2\over 2(r^2-\ell^2)}\in[-1, 3/2].
\end{equation}

To find the critical point where a Bardeen RBH switches over to a Bardeen ulta-compact object you need to simultaneously solve
\begin{equation}
    r= 2m(r); 
    \qquad \hbox{and} \qquad 
    1 = 2m'(r).
\end{equation}
This happens when
\begin{equation}
    r^2 = 2 \ell^2; \qquad \qquad 
    \ell = {4\sqrt{3}\over 9} M. 
\end{equation}
So there are definitely many qualitative similarities to the Hayward case (for instance, we still have $w_t\approx -1$ in the de Sitter core $r \ll \ell$); but a few technical (quantitative) changes.


\section{Unifying the Hayward and Bardeen metrics}

\noindent
To unify the discussion of the Hayward and Bardeen metrics we could (setting $\Phi(r)=0$) consider a quasi-local mass function $m(r)$ of the form:
\begin{equation}
m(r)=\frac{Mr^3}{\left(r^{(2+n)}+\ell^{2} (2M)^{n}\right)^{(3-n)/2}}. 
\end{equation}
Hayward is $n=1$ and Bardeen is $n=0$. In this framework one would find it convenient to set $\bar\ell = (\ell^2 M^n)^{1/(2+n)}
= M (\ell/M)^{2/(2+n)}$.

\noindent
Another possibility would be to take
\begin{equation}
m(r) = {M r^3 \over (r^n + [ \ell^a M^{(1-a)}]^n )^{(3/n)} }.    
\end{equation}
Bardeen would be $n=2$; $a= 1$,
Hayward would be $n=3$; $a=2/3$. But this form of the quasi-local mass is a little trickier to use. In this framework one would find it convenient to set $\bar\ell = \ell^a M^{(1-a)}= M(\ell/M)^a $.

\noindent
Overall, we see that attempting to generalize the discussion from Hayward to Bardeen geometries, while introducing several quantitative technical changes, does not really involve any qualitatively new concepts.

\section{Light rings and trapped regions}\label{sec:LR}
In Sec.~\ref{eq:haywex}, we have seen that the Hayward geometry has only one light ring for the values of $\alpha$ for which a trapped region is present, and two light rings when there is no trapping region. In the limiting case in which $\alpha=\alpha_\star$, the stable light ring coincides with the degenerate horizon. Here we show that this structure is completely general and holds true for any spherically symmetric geometry with a trapped region. 

In $(t,r,\theta,\phi)$ coordinates, the location of the light rings can be obtained finding the local extrema of the effective potentials \cite{Cunha:2017qtt,Cunha:2022gde},
\begin{equation}
    V_\pm=\frac{g_{t\phi}\pm\sqrt{g_{t\phi}^2-g_{tt}g_{\phi\phi}}}{g_{\phi\phi}}\,.
\end{equation}
For a spherically symmetric configuration 
\begin{equation}
   ds^2 =-e^{-2\Phi(r)}F(r)\text{d}t^2
    +F^{-1}(r)\,\text{d}r^2+r^2\text{d}\Omega^2\,,
\end{equation}
the condition for the light rings reads
\begin{equation}
    \partial_r V=0\qquad\iff\qquad\frac{d}{dr}\left(\frac{e^{-\Phi(r)}}{r}\sqrt{-F(r)}\right)=0\,.
\end{equation}
Which is equivalent to 
\begin{equation}\label{eq:LRs}
    g(r):=\frac{1}{2}F'(r)-\left( \Phi'(r)+\frac{1}{r}\right)F(r)=0\,,
\end{equation}
where the prime denotes derivative with respect to $r$. This equation is guaranteed to have an even number of solutions (counting multiplicity) \cite{Cunha:2017qtt}. However, such solutions corresponds to light rings only if $F(r)>0$, as otherwise the effective potential would be imaginary. This implies that any solution of Eq.~\eqref{eq:LRs} inside a trapped region must be discarded, so that the corresponding spacetime has an odd number of light rings. 
It is easy to show that there is an odd number of light rings whenever a trapped region is present. In fact, both at the inner and outer horizons we have $F=0$, while we have $F'<0$ at the inner horizon and $F'>0$ at the outer horizon. Therefore,
\begin{equation}
    g(r_-)<0\,,\,\qquad g(r_+)>0\,.
\end{equation}
This shows that there is always an odd number of roots of Eq.~\eqref{eq:LRs} between the two horizons. Therefore, a regular black hole has an odd numbers of light rings, and we can conclude that the structure of Fig.~\ref{fig:HLR} is general for every spherically symmetric static spacetime with a trapped region. 

\section*{Acknowledments}

RCR acknowledges financial support through a research grant (29405) from VILLUM fonden.
FDF acknowledges financial support by Japan Society for the Promotion of Science Grants-in-Aid for international research fellow No. 21P21318. 
SL acknowledges funding from the Italian Ministry of Education and  Scientific Research (MIUR)  under the grant  PRIN MIUR 2017-MB8AEZ. 
CP acknowledges the financial support provided under the European Union's H2020 ERC, Starting Grant agreement no.~DarkGRA--757480 and support under the MIUR PRIN and FARE programmes (GW- NEXT, CUP: B84I20000100001).
MV was supported by the Marsden Fund, via a grant administered by the Royal Society of New Zealand. 

\bigskip
\bibliographystyle{unsrt}

\bibliography{refs}

\bigskip
\bigskip
\hrule\hrule\hrule
\end{document}